\documentstyle[12pt]{article}
\begin{document}
\setlength{\baselineskip}{.375in}
\vspace{1in}
\begin{center}
\Large\bf
Behaviour of the energy gap\\
near a\\
commensurate-incommensurate transition\\
in\\
double layer quantum Hall systems at $\nu=1$\\
\vspace{0.75in}
\large\sc N. Read\\
\vspace{0.325in}
\normalsize\em
Departments of Physics and Applied Physics, P.O. Box 208284\\
Yale University, New Haven, CT 06520\\
\vspace{0.75in}
\end{center}
\rm
The charged excitations in the system of the title are vortex-antivortex pairs
in the spin-texture described in the theory by Yang et al which,
in the commensurate phase, are bound together
by a ``string''. It is shown
that their excitation energy drops as the string lengthens as the parallel
magnetic field approaches the critical value, then goes up again in the
incommensurate phase. This produces a sharp downward cusp at the critical
point. An alternative description based on the role of disorder in the
tunnelling and which appears not to produce a minimum in the
excitation energy is also discussed. It is suggested that a similar
transition could also occur in compressible Fermi-liquid-like states.

\begin{verbatim}
PACS Nos. 73.40.Hm, 64.70.Rh, 73.20.Dx
January, 1995
\end{verbatim}
\newpage

Recent experiments on double layer quantum Hall systems \cite{murph},
consisting of two two-dimensional electron gases (2DEGs) separated by only
about $d\simeq 100$ \AA\ in a double quantum well structure, have clarified
the phase diagram in a perpendicular magnetic field $B$ at Landau level
filling factor $\nu=1$ and, surprisingly, showed evidence of an apparent
phase transition, in that the activation energy gap drops dramatically when
a parallel magnetic field is introduced, saturating at a constant value for
higher $B_\parallel$. A theory for a phase transition for such a system was
proposed \cite{yang}, but in that work the activation gap itself was not
discussed. The present paper is mainly concerned with this issue. The Indiana
group has independently reached some of the same conclusions in more recent
work \cite{moon}. I also discuss possible effects of disorder.

The theory \cite{yang,moon}, which draws upon earlier work
\cite{wenzee,halpetc,leekane,sondhi,ezawa}, can be
summarised as follows. I will
not discuss the SU(2) symmetry that is present only in the limit
$d\rightarrow0$ which is physically unrealizable in this context. All we
need to know is that when the total filling factor $\nu=n\Phi_0/B$ is of the
form $1/q$, $q$ odd (and presumably also at some hierarchical generalizations,
in particular whenever the corresponding system with real spin is spin
polarized in the absence of Zeeman splitting) the system (at $B_\parallel=0$)
forms an incompressible fluid exhibiting a quantum Hall plateau, at least for
$d$ not too large (for larger $d$, the layers can form essentially independent
even denominator compressible states \cite{HLR}). The density of electrons in
each layer, labelled $\uparrow$, $\downarrow$, is essentially the same, $=n/2$.
I denote the density difference $\varpi({\bf r})={\frac{1}{2}}(n_\uparrow({\bf
r})-n_\downarrow({\bf r}))$ (pronounced {\it pi}). It turns out that, in the
states under discussion, there is long range order in the phase $\theta$
canonically conjugate to $\varpi$, so that $[\theta({\bf r}),\varpi({\bf r}')]=
i\delta({\bf r}-{\bf r}')$ or more accurately since $\theta$ is periodic we
should use $e^{i\theta}$ (with complex eigenvalues of modulus unity) and
$[\varpi({\bf  r}),e^{i\theta}({\bf r}')]=e^{i\theta}({\bf r})\delta({\bf
r}-{\bf r}')$. This operator can be viewed as $e^{i\theta}({\bf r})
\simeq 2n^{-1}c_\uparrow^\dagger({\bf r})c_\downarrow({\bf r})$ which
transfers an electron at ${\bf r}=(x,y)$ from the lower to the upper layer.
The long range phase coherence of this operator is a consequence of interlayer
correlations (two electrons avoid each other whether they are in the same layer
or not) induced by interactions.

Since charge fluctuations (in $n_\uparrow({\bf r})+n_\downarrow({\bf r})$) are
gapped, the Hamiltonian for the system including the tunnelling term with
coefficient $t$, and in the presence of the parallel magnetic field
$B_\parallel$ in the $y$ direction, can be written in terms of the ``isospin''
variables only, as
\begin{equation}
H=\int d^2r\,\left\{\frac{1}{2}\chi^{-1}\varpi^2 +
\frac{1}{2}\rho_s|\nabla\theta|^2 - t\cos(\theta-Qx)\right\}.
\label{ham}
\end{equation}
The coefficients deserve comment: $Q=2\pi B_\parallel d/\Phi_0$ \cite{yang},
$t$ is related to Murphy et al's tunnelling gap $\Delta_{SAS}$ by
$t\simeq\nu\Delta_{SAS}/4\pi \ell_B^2$ and (for $\nu=1$) to
Yang et al's $t$ by $t\rightarrow t/2\pi \ell_B^2$ ($\ell_B$ is the magnetic
length $2\pi\ell_B^2=\Phi_0/B$).
$\rho_s$ is the ``isospin stiffness'' resulting from the change in exchange
energy when $\theta$ varies, and $\chi$ is a charge susceptibility related to
the capacitance per unit area (this term is omitted in most equations in
\cite{yang}, but appears as $U_z$ in their eq.\ (4)). As $d\rightarrow0$ (the
``SU(2) limit''),
$\chi$ diverges as $d^{-1}$, $\rho_s\propto e^2/\epsilon\ell_B$ can be
calculated exactly for $\nu=1$ \cite{sondhi}, and $t=\nu\Delta_{SAS}/4\pi
\ell_B^2$; all these quantities are modified at finite $d$.

One other fact is important: I allow states where $e^{i\theta}$ becomes
undefined at isolated points, and winds around the unit circle once or more (in
either sense) around such points, so there is a vortex at these points. The
underlying theory predicts that a vortex carries a real electric charge
corresponding to $\pm\frac{1}{2q}$ of an electron for a $2\pi$ vortex in
$\theta$;
{\em the charge can take either sign, independent of the sign of the vortex}.
Therefore the Hamiltonian (\ref{ham}) is implicitly supplemented with a
Coulomb interaction between the charges on the vortices. I will consider only
the smallest vortices, of magnitude $\pm 2\pi$. They have quantum mechanical
dynamics which can be viewed (at zeroth approximation, neglecting changes in
the
densities $n_\uparrow$, $n_\downarrow$ from their averages = $\frac{1}{2}n$) as
that of charges $\pm\frac{1}{2q}e$ in the same external field $B$ and
restricted
to their lowest Landau level; thus they will drift perpendicular to any
electric field they experience. I specialize to $q=1$ from
now on.

To understand the phase transition as $B_\parallel$ (or $Q$) is increased, it
is helpful to consider the imaginary time path integral representation of the
partition function, as $\beta=1/k_BT\rightarrow\infty$:
\begin{eqnarray}
Z&=&{\rm Tr}\,e^{-\beta H}\\
 &=&\int {\cal D}[\theta]\,e^{-S}
\end{eqnarray}
where
\begin{equation}
S=\int d^2 r\int_0^\beta d\tau\,\left\{\frac{1}{2}\chi\left({\partial\theta
\over
\partial\tau}\right)^2 + \frac{1}{2}\rho_s|\nabla\theta|^2
-t\cos(\theta-Qx)\right\}.
\label{action}
\end{equation}
As is common, the $T=0$ quantum transition in $D=2$ space dimensions is
equivalent to a classical finite temperature transition in $D+1=3$ dimensions
(the temperature in the classical problem has been absorbed into the
parameters). The classical problem has been well studied, and we can make use
of existing results, reviewed in \cite{bak,dennijs}.

A first step is to solve the problem classically (as $\hbar\rightarrow 0$) by
simply minimizing $S$ or equivalently $H$ regarded as a
classical Hamiltonian. It is convenient to define a new phase
$\theta'=\theta-Qx$; this corresponds to a change of gauge in the original
problem. In terms of $\theta'$, $S$ is
\begin{equation}
S=\int d^2 r\int_0^\beta d\tau\,\left\{\frac{1}{2}\chi\left({\partial\theta'
\over
\partial\tau}\right)^2 + \frac{1}{2}\rho_s\left[\left({\partial\theta'
\over
\partial y}\right)^2
+
\left({\partial\theta'
\over
\partial x}+Q\right)^2\right]
-t\cos\theta'\right\}.
\label{action'}
\end{equation}
The minimum will have $\theta'$ independent of $\tau$, $y$. $Q$ enters through
a total derivative which
does not affect the equation obtained by varying $\theta'$ which is
\begin{equation}
\rho_s{d^2\theta' \over dx^2}=t\sin\theta'.
\end{equation}
This equation has well known soliton solutions in which $\theta'$ changes by
$2\pi$ at each domain wall of width $\sim(\rho_s/t)^{1/2}$. The role of $Q$ is
to act as a chemical potential for domain walls. For $Q\leq
Q_c=8(t/\rho_s)^{1/2}/2\pi$ \cite{frank,bak} the minimum action $S$ is obtained
by selecting the
solution $\theta'=0$, with no domain walls. This is the commensurate phase, in
which the tunnelling $t$ overcomes the stiffness $\rho_s$, because
$\partial \theta/\partial x$ is small. When the chemical potential exceeds the
action per unit area (in the $y$, $\tau$ plane) of a wall, i.e.\ when
$Q>Q_c$, it is advantageous to form walls. The walls repel each other
exponentially due to the exponential relaxation of $\theta'$ to $0$ (mod
$2\pi$) outside the
``core'' region of each wall. Consequently the minimum action is obtained for a
periodic solution (the ``striped incommensurate solid'') in which the number
of walls per unit length in the $x$ direction,
$\bar{\ell}^{-1}\equiv\bar{Q}/2\pi$,
is \cite{frank,bak}
\begin{equation}
\bar{Q} \sim 1/\ln[1/(Q-Q_c)].
\label{Qbar}
\end{equation}
This wall spacing $\bar{\ell}$ defines the correlation length
$\xi_x=\bar{\ell}$ in the incommensurate phase,
and diverges only logarithmically as $Q\searrow Q_c$, thus the critical
exponent $\nu$ (not to be confused with the filling factor) is zero. For
illustrations of a wall configuration and of $\bar{Q}$ versus $Q$, see Figs 11,
12 in \cite{bak}. A useful variable in the incommensurate phase is
\begin{equation}
\bar{\theta}\equiv \theta'+\bar{Q}x.
\end{equation}
In the periodic array of domain walls, $\bar{\theta}$ is then constant on
length scales $>2\pi/\bar{Q}$. As $Q\rightarrow\infty$, $\bar{Q}\rightarrow Q$
and so $\bar{\theta}\rightarrow \theta$. In this limit, $\theta$ has ceased to
rotate to follow $Qx$, even on short length scales; it is as if $t$ went to
zero.

Next we must consider the role of fluctuations. For the case $D=1$, Pokrovsky
and Talapov \cite{poktal,bak,dennijs} showed that quantum fluctuations (or
thermal fluctuations in the equivalent 2 dimensional classical problem) lead to
$\bar{Q}\sim (Q-Q_c)^{1/2}$ instead of (\ref{Qbar}). In general dimension $D$,
the walls are $D$-dimensional objects, and are effectively flat for $D>2$,
wander due to fluctuations for $D<2$, and for $D=2$ are only logarithmically
rough. Fisher and Fisher \cite{fishfish} argued that in the latter case, this
roughening is unimportant and (\ref{Qbar}) still holds (the application to the
present $T=0$ $D$-dimensional quantum situation is mentioned in their paper).
It is also argued that walls that end on a vortex line are not important, and
we
expect that the same is true here, even though the action of the vortex lines
differs due to their different (lowest Landau level) dynamics.

Now consider the case where the physical temperature $T$ is nonzero. Apart from
some renormalization of parameters at low $T$ (which will be unimportant
because fluctuations of the walls at $T=0$ are unimportant), the problem is
described by the
same form of ``action'' (now classical Hamiltonian) given by (\ref{action}) or
(\ref{action'}) with the $\tau$ dependence of
$\theta$ or $\theta'$ dropped and so $\int d\tau$ replaced by a factor $\beta$.
As it stands, this model has a transition of the
Pokrovsky-Talapov type. However, it is known that, because of vortices that
terminate a domain wall and hence act as dislocations in the domain wall
``solid'', the finite $T$ transition is preempted by a Kosterlitz-Thouless (KT)
transition to an incommensurate fluid as $Q_c$ is approached from above,
because this is the
$p=1$ case of the more general theory \cite{coppersmith}. Note that at finite
$T$, $\langle e^{i\bar{\theta}({\bf r})} e^{-i\bar{\theta}({\bf 0})} \rangle
\sim r^{-\eta(T)}$ in the ``solid'' phase, but $\sim e^{-r/\xi}$ in the fluid,
for some correlation length $\xi$. Also $\langle e^{i\theta'({\bf r})}
e^{-i\theta'({\bf 0})} \rangle \sim e^{i\bar{Q}x} e^{-r/\xi}$ in the
incommensurate fluid, but as $Q$ is decreased further, there is in principle a
boundary where $\bar{Q}$ reaches zero and stays there for smaller $Q$ (the
``commensurate fluid''). This is not a true transition since no broken
symmetries are involved, and $\xi$ remains finite as it is
crossed. Further, its position may depend on details of its definition
\cite{dennijs}. In the $(Q,T)$ plane, both this line and the KT transition line
approach $Q=Q_c$ as $T\rightarrow 0$. In the present case of
$p=1$, there is no other finite $T$ transition between the
$\bar{Q}\rightarrow0$ line and the zero temperature axis. Note that the
KT transition extends all the way in to $B_\parallel=0$ when $t\rightarrow 0$;
the transition temperature $T_c$ goes to a constant at large $Q\gg Q_c$ for
finite $t$ which is the same as the value at all $Q$ when $t=0$
\cite{yangpriv}.

After the above, somewhat lengthy, review of earlier work, we turn to the
question of how the energy gap of the charged excitations, as seen
experimentally in the thermally activated longitudinal resistance, behaves as
$Q$ increases and passes through $Q_c$. The basic excitations are vortices of
charge $\pm\frac{1}{2}e$. Since $\theta$ winds by $2\pi$ around the vortex, for
$t\neq0$ a vortex is at the end of a domain wall in spacetime (or ``string'' in
space), which when $Q$ is zero has an
energy $\sim(\rho_s t)^{1/2}$ per unit length in the $x$-$y$ plane (I again
begin by considering the classical $\hbar=0$, $T=0$ system, and taking
$\tau$-independent configurations). This string tension means that
the domain wall must end at another vortex of opposite vorticity in order to
give finite free energy. Thus vortices will be ``confined'' by the linearly
increasing potential between opposite vortices; no free vortices exist. We are
interested in such vortex pairs with total charge $\pm e$, so the charges at
the ends repel each other. (The other case, where the charges sum to zero,
gives an excitation that should be identified with the usual gapped collective
mode.) For nonzero $Q$, the string tension depends on the orientation of the
wall and is smallest when it is parallel to the $y$ axis (and largest when it
is
antiparallel). As $Q$ increases, the string
tension decreases, the string
lengthens, and the minimum excitation energy $E_{\rm min}$ decreases.
At $Q=Q_c$ the
string tension is zero for a string parallel to the $y$ axis, and is negative
for $Q>Q_c$ so infinitely long domain walls condense spontaneously to form the
``striped incommensurate solid''.
The total energy of the bound vortex-antivortex pair for $Q<Q_c$, taking
a straight string of length $R$ and of orientation angle $\phi$ relative
to the $y$ axis, and including a ``core energy'' for each vortex, is thus
\begin{equation}
E(R,\phi)=2E_{\rm core}+{e^2\over 4\epsilon R}+2\pi\rho_s R(Q_c-Q\cos\phi).
\label{E(R,phi)}
\end{equation}
The minimum at fixed $\phi$ is
\begin{eqnarray}
R(\phi)&=&[e^2/8\pi\epsilon\rho_s(Q_c-Q\cos\phi)]^{1/2}\\
E_{\rm min}(\phi)&=&2E_{\rm core}
              +2[2\pi\rho_s(Q_c-Q\cos\phi)e^2/4\epsilon]^{1/2}.
\label{gaphi}
\end{eqnarray}
The minimum is then $E_{\rm min}(0)$. At $Q=Q_c$, $E_{\rm min}(0)
=2E_{\rm core}$.
These expressions are of course accurate when $R$ is large compared with the
width of the wall, $\sim (\rho_s/t)^{1/2}$, that is when
\begin{equation}
{e^2t\over 8\pi\epsilon\rho_s^2(Q_c-Q\cos\phi)}\gg O(1).
\label{criterion}
\end{equation}
In the experimental context, the long wavelength forms (\ref{ham}),
(\ref{action}) are valid when
$(\rho_s/t)^{1/2}\gg\ell_B$, which, since $\rho_s\sim e^2/\epsilon\ell_B$
and assuming finite $d$ corrections are not too large, is equivalent to
$\Delta_{SAS}/(e^2/\epsilon\ell_B)\ll 1$, which is easily satisfied; however
the criterion (\ref{criterion}) becomes, for $Q=0$,
$\Delta_{SAS}/(e^2/\epsilon\ell_B)\gg 1$ which is not compatible. However, it
may still be that for small $Q$ there is a tendency for the charge $\pm e$
excitation to separate slightly into two lumps of charge $\pm\frac{1}{2}e$;
again this is a topic for numerical study. On the other hand, as $Q$ increases
the above considerations become more accurate.
Once the string picture is applicable, the experiment will actually pick up
contributions from all orientations (and lengths) of the string, weighted by a
Boltzmann factor involving $E(R,\phi)$; this will be
especially important at small $B_\parallel$ where the variation of the gap
with $\phi$ becomes smaller than $k_B T$ for accessible $T$. Thus the
measured ``gap'' will {\em not} be given simply by $E_{\rm min}(0)$, which
varies linearly with $Q$ at small $Q/Q_c$, but will have a much flatter
appearance, $\sim Q^2$. Near $Q=Q_c$, however, the form $E_{\rm min}(0)$
should dominate. Note that in this region there are still excited states of
the bound pair, with constant density of states just above $E_{\rm min}(0)$;
this will contribute a prefactor $\sim T$ to the exponentially activated
temperature dependence.

The preceding paragraph was the classical picture, which will be useful when
the temperature is higher than the quantum-mechanical energy splittings. For
the quantum dynamics of the bound pair, within the straight string
approximation, we should recall that each vortex behaves as a charge
$\pm\frac{1}{2}e$ particle in the field $B$, restricted to its lowest
Landau level. (The quantum mechanics of a similar, but alas not identical,
system including a flexible string has been studied in \cite{callan}.)
The bound pair system then behaves as a charge $\pm e$ particle in
the lowest Landau level with the potential $E(R,\phi)$. For a first look, we
consider the semiclassical approximation. The particle moves on the
equipotentials of $E(R,\phi)$ but is subject to the quantization condition that
an orbit encloses an integral number of flux quanta. The $Q=0$ case is simple
due to its rotational invariance; the quantization condition is $\pi R^2=2\pi m
\ell_B^2$, where the integer $m$ is essentially the angular momentum.
If this condition is satisfied with $m=$ some $m^*$ at
the minimum of $E(R,\phi)$ at $R^*=R(\phi)$ as given above, then using a
quadratic approximation to the radial potential near $R=R^*$ we find that the
energy difference of the $m=m^*$ and $m=m^*\pm1$ states is
\begin{eqnarray}
\Delta E&=&{e^2\ell_B^4\over\epsilon {R^*}^5}  \nonumber \\
        &=&2^{25/2}\ell_B^4 (\rho_s t)^{5/4}(\epsilon/e^2)^{3/2}.
\label{quenergysplit}
\end{eqnarray}
As $R(\phi)$ varies (at $Q=0$, by varying e.g.\ the density of electrons),
levels will cross, since each level has different angular momentum, so the
energy gap will actually exhibit quadratic minima separated by cusps, on
top of the classical behaviour $E_{\rm min}(\phi)$. The scale of this variation
will be given by $1/4$ of (\ref{quenergysplit}). At $Q\neq 0$, angular
momentum is
no longer conserved. For small $Q$ we may use perturbation theory around the
$Q=0$ limit to calculate $E_{\rm min}$. Since $Q$ enters the Hamiltonian
through $R\cos\phi$, it changes angular momentum by $\pm 1$ so the first
correction to $E_{\rm min}$ is of order $Q^2$ (this statement
does not depend on the use of the string and semiclassical
approximations). As $Q$ varies,
or as other parameters vary at nonzero $Q$, the ground state energy will
now exhibit variations similar to those at $Q=0$, but level crossings
will be avoided.
However such structure will be washed out as $Q$ increases toward $Q_c$.
Semiclassically, it is clear that as $Q$ increases from zero, a point is
reached at which the lowest allowed orbit changes topology from encircling the
origin to not encircling the origin. For larger $Q$, the variation of $E_{\rm
min}$ with $Q$ will have no oscillating component. Thus it is clear that
quantum effects in $E_{\rm min}$ are not important as $Q\rightarrow Q_c$.

Turning to the incommensurate phase, for $Q$ not too much larger than $Q_c$ so
we can
still use the picture of well-separated domain walls, we might imagine that
free vortices can exist, since domain walls have condensed anyway and can
simply readjust to accomodate the wall ending at the vortex. However, this
introduces distortion into the array of walls, equivalent to a dislocation.
The elastic energy will cause the usual logarithmic divergence in the total
energy of such a configuration. Consequently, vortices are again bound in
pairs, but the attractive term is now logarithmic for large separation
instead of linear. To calculate the behaviour of the gap, we need the elastic
constants of the domain wall solid. I have been unable to find such a
calculation for this $T=0$ case in the literature, so it is worked out below.

Once again our main interest is in time-independent configurations, whose
long-wave\-length potential energy can be expressed in terms of $\bar{\theta}$
as
(neglecting an additive constant)
\begin{equation}
E=\int d^2r\,\frac{1}{2}\left(K_x\left({\partial\bar\theta\over\partial
x}\right)^2 + K_y\left({\partial\bar\theta\over\partial
y}\right)^2 \right).
\label{arrayen}
\end{equation}
In terms of domain walls, it is evident that nonzero
$\partial\bar{\theta}/\partial x$
represents a change in the separation of the walls, while nonzero
$\partial\bar{\theta}/\partial y$ represents a rotation of the array away from
its preferred orientation with walls parallel to the $y$ axis. To obtain the
energies, we consider the energy of a periodic array of straight walls,
described by $\mbox{\boldmath $\ell$}=(\ell_x,\ell_y)$, the vector
center-to-center
separation of adjacent walls; the ground state will be
$\mbox{\boldmath $\ell$}=(\bar{\ell},0)$. On scales
$\gg \ell=|\mbox{\boldmath $\ell$}|$,
we have
$\nabla\theta'=-2\pi\mbox{\boldmath $\ell$}/\ell^2$ (note the minus sign due to
our definitions in (\ref{action'})) and hence for small deviations from the
ground state $\nabla\bar{\theta}\simeq
2\pi\bar{\ell}^{-2}(\bar{\ell}-\ell_x,-\ell_y)$. Generalizing a formula from
\cite{bak,fishfish}, the energy density is
\begin{equation}
E/L^2=-2\pi\rho_s Q\ell_x/\ell^2 + 2\pi\rho_s Q_c/\ell
             +We^{-\kappa\ell}/\ell.
\label{incommenergy}
\end{equation}
$We^{-\kappa\ell}$ is the interaction energy per
unit length of the array of walls; $\kappa$ is exactly $(t/\rho_s)^{1/2}$ in
the classical approximation, and $W=32(\rho_s t)^{1/2}$ \cite{bak}.
The energy density is minimized for $\bar{\ell}$ given by
\begin{eqnarray}
\kappa\bar{\ell}e^{-\kappa\bar{\ell}}&\sim&2\pi(Q-Q_c)/32\kappa\nonumber\\
                 \bar{\ell}&\sim&
            (\rho_s/t)^{1/2}\ln(32(t/\rho_s)^{1/2}/2\pi(Q-Q_c)),
\end{eqnarray}
for $\kappa\bar{\ell}\gg 1$, as stated earlier.
Expanding the energy density near its minimum at $(\bar{\ell},0)$ we obtain
\begin{eqnarray}
K_x&=&(\rho_s t)^{1/2}(Q-Q_c)\bar{\ell}^2/2\pi\nonumber\\
K_y&=&\rho_s Q_c\bar{\ell}/2\pi=8(\rho_s t)^{1/2}\bar{\ell}/(2\pi)^2.
\label{Ks}
\end{eqnarray}
Thus near the transition, the array of walls is much softer in the
direction perpendicular to the walls. When the walls are no longer dilute,
however, the constants $K_x$, $K_y$ will $\rightarrow \rho_s$
\cite{yangpriv}. Note that from (\ref{incommenergy}), (\ref{Ks}) we can obtain
the behaviour of the finite $T$ KT transition temperature, $T_c\sim
(K_xK_y)^{1/2}\sim (Q-Q_c)^{1/2}\ln^{3/2}[1/(Q-Q_c)]$.

For $Q>Q_c$, the energy of a well-separated vortex-antivortex pair of total
charge $\pm 1$ will be, for separation $X$, $Y$, $R^2=X^2+Y^2$,
\begin{equation}
E(X,Y)=2E_{\rm core}+{e^2\over 4\epsilon R}+C'(K_xK_y)^{1/2}\ln\{(K_y X^2+K_x
Y^2)/K_y\bar{\ell}^2\}^{1/2}
\label{Einc}
\end{equation}
where $C'$ is another constant, the additional core energy due to the domain
wall solid is neglected. The logarithmic term was obtained by first rescaling
the coordinates to make the energy isotropic, and using the wall separation as
the lower cutoff on the $X$ separation at which the logarithm is arbitrarily
set to zero. The logarithm is meaningful only at larger separations, which for
the $Y$ direction means $Y>(K_y/K_x)^{1/2}\bar{\ell}$.
Minimizing once again, it is found that the energy is lowest when the pair is
oriented in the $y$ direction and is then given by $E_{\rm
min}=2E_{\rm core}+C'(K_xK_y)^{1/2}(1+\ln(e^2/4\epsilon C'K_y\bar{\ell}))$.
However the corresponding value of $Y$ is $Y=e^2/(4\epsilon C'(K_xK_y)^{1/2})$
which  as $Q\rightarrow Q_c$, is smaller than the condition for the validity
of the logarithmic interaction by a factor $\sim \ln^2[1/(Q-Q_c)]$.
Therefore, the form (\ref{Einc}) is not, in fact, appropriate.

To try to obtain the correct asymptotic form of the energy gap near $Q_c$, let
us consider the opposite limit, where the separation of the vortex and
antivortex is in the $y$ direction and is much less than the length
$(K_y/K_x)^{1/2}\bar{\ell}$. There are two cases to consider: the sign
of the separation is such that either ({\it
i\/}) an additional short
length of parallel string is inserted between the walls in the array, or ({\it
ii\/}) a short length of a wall is removed from the array. Here ``short'' means
compared with $(K_y/K_x)^{1/2}\bar{\ell}$, but not necessarily compared with
$\bar{\ell}$. At such short separations, the distortion of the array may become
small, that is the effective dipole moment of the pair, as determined by the
$\bar{\theta}$ field far from the pair, may be much less than their actual
moment, as defined by the vorticity times the separation. Even if the elastic
contribution to the vortex-antivortex potential
is not zero, there is still likely to be a linear piece, as will be described
shortly, that dominates the potential, and is not screened out as it will be
for large separations by elastic deformation of the array that produces the
asymptotically logarithmic behaviour.
For case ({\it i\/}), the energy of the pair is then
\begin{eqnarray}
E(Y)&=&2E_{\rm core}+{e^2\over 4\epsilon Y} + 2\pi\rho_s(Q_c-Q)Y
                        + 2W e^{-\kappa\bar{\ell}/2} Y \nonumber\\
    &\simeq&2E_{\rm core}+{e^2\over 4\epsilon Y}
+2(2\pi\rho_sW(Q-Q_c)/\kappa\bar{\ell})^{1/2}Y
\label{Ecorr}
\end{eqnarray}
so the minimum is found to be
\begin{eqnarray}
Y&=&[e^4\kappa\bar{\ell}/128\pi^2\epsilon^2\rho_s^2W^2(Q-Q_c)]^{1/4}
\nonumber\\
E_{\rm min}&=&2E_{\rm core}+
2[2\pi\rho_sW(Q-Q_c)e^4/4\kappa\bar{\ell}\epsilon^2]^{1/4}.
\label{Emininc}
\end{eqnarray}
For case ({\it ii\/}), the pair have the opposite orientation from case ({\it
i\/}), but in terms of the magnitude $Y$ of the separation, the energy is now
\begin{eqnarray}
E(Y)&=&2E_{\rm core}+{e^2\over 4\epsilon Y} + 2\pi\rho_s(Q-Q_c)Y
                        - 2W e^{-\kappa\bar{\ell}} Y \nonumber\\
    &\simeq&2E_{\rm core}+{e^2\over 4\epsilon Y}+ 2\pi\rho_s(Q-Q_c)Y
          - 4\pi\rho_s(Q-Q_c)Y/\kappa\bar{\ell}
\label{Ecorr'}
\end{eqnarray}
Note the opposite signs from case ({\it i\/}), since a length of wall has been
removed rather than added, and that the interaction term containing $W$ has the
same form as in (\ref{arrayen}), while in case ({\it i\/}) the argument of the
exponential was
halved because the inserted walls were only $\bar{\ell}/2$ away from the
others. This term in (\ref{Ecorr'}) can now be neglected, and the form for
the minimum is therefore similar to (\ref{gaphi}). Since the resulting $E_{\rm
min}$ has a square-root dependence, this is asymptotically lower than
(\ref{Emininc}), allowing us to write a
single unified formula for our final result on either side of the transition,
\begin{eqnarray}
Y^*&=&\left({e^2 \over 8\pi\epsilon\rho_s|Q-Q_c|}\right)^{1/2}\nonumber\\
E_{\rm min}&=&2E_{\rm core}+(2\pi e^2\rho_s|Q-Q_c|/\epsilon)^{1/2}.
\label{final}
\end{eqnarray}
Thus there is a symmetrical downward square-root cusp in $E_{\rm min}$ at
$Q_c$, with no direct dependence on the tunnelling $t$.
The value of $Y^*$ is smaller by a factor
$\bar{\ell}^{-1/2}\sim\ln^{-1/2}1/(Q-Q_c)$ than
$(K_y/K_x)^{1/2}\bar{\ell}$ as $Q\rightarrow Q_c$ on the incommensurate side,
so the calculation should be asymptotically valid. Quantum corrections should
be asymptotically negligible on both sides.

Since the binding of the
vortices will be somewhat weaker than at $Q=0$ even when $Q\gg Q_c$, due to the
effective removal of $t$ from the action (\ref{action}), it is to be
expected that the activation gap saturates at a value below its
$B_\parallel=0$ value at large $B_\parallel$ \cite{yangpriv}, as is observed
very strikingly in the experiments \cite{murph}.

So far the cusp in the activation gap has not been observed in experiments.
While this might be due simply to large scale variations in the density and
hence in the value of $Q_c$, another possibility is that it is connected with
the effects of small scale disorder. M.P.A. Fisher \cite{fisher} has pointed
out that randomness in the tunnelling $t$ is a relevant perturbation in the
incommensurate phase, which can destroy the ordered state both at finite and
at zero temperature. If the tunnelling varies from place to place, perhaps
because of defects such as steps in the layer separation, then $t$ in eq.\
(\ref{ham}) varies with $x$, though it probably remains real. In the
incommensurate phase, in the $\bar{\theta}$ variables that describe the array
at scales large compared with the spacing, the randomness in $t$ causes
deviations in the array of walls that can still be described in terms of a term
like $-t({\bf r})\cos(\bar{\theta}({\bf r}))$. It is well-known that such a
term
(which tends to pin the domain walls randomly) destroys the long-range order
(or quasi-long-range order at $T\neq0$) whenever
the spacetime dimension is less than $4$, no matter how weak the disorder.
While
the magnitude and correlation length of the disorder in the samples are
uncertain, it is worthwhile working out the consequences of this effect for the
theory of the activation gap. At large enough length scales the disorder will
destroy the ordering, and this should affect the behaviour of the charged
excitations which as we have seen become very large near the transition.

In the presence of disorder, I claim that a zero temperature phase transition
can still occur as $B_\parallel$ (or $Q$) is increased, since the state at
$Q=0$ is essentially stable to weak disorder, which can at worst only
introduce short lengths or loops of string into the ground state described for
the pure system. At some $Q_c$ it becomes favourable to condense infinitely
long domain walls, but they
will now form a randomly pinned array that may contain dislocations (vortices)
due to the quenched disorder. Since there is no long-range order, there can be
no asymptotically logarithmic potential to confine the charged vortex
excitations into pairs, and so the possibility of $\pm e/2$ excitations exists.
Accordingly, it is plausible that the energy gap decreases as $Q$ increases to
$Q_c$, and then saturates at an excitation energy $E_{\rm core}$ immediately
$Q>Q_c$.

The critical properties of the transition itself will of course also be
affected by the destruction of the ordered state. Since the time-independent
disorder distinguishes the imaginary time direction from the two space
directions, it is possible that correlations scale at criticality with
different powers of separation in each of the three directions ($x$, $y$,
$\tau$).
It is also possible that the disorder effectively restores the isotropy of
the scaling in the two space directions; an analogous phenomenon occurs
in the pure system at the finite temperature KT transition. This would
leave a single dynamic exponent $z$ relating distances and times,
$\tau\sim r^z$. All of these properties of this quantum phase transition
are completely open at present. It would be interesting to study them by
Monte Carlo simulation in the simple model consisting of eq.\ (\ref{ham})
(on a lattice) plus disorder, neglecting the subtle effects of the lowest
Landau level dynamics of the vortices.

The theory presented above extends to other fractions $\nu$, so long as
an incompressible ground state that is fully polarized in the isospin
variable $e^{i\theta}$ is expected at $B_\parallel=0$. It is also interesting
to consider compressible states of the type discussed in \cite{HLR}. In
particular, for $\nu=1/q$, $q$ even, a compressible Fermi-liquid-like state is
possible at short separations $d$ \cite{halpnewport}, in which the
Laughlin-Jastrow type correlation factor is independent of which layer the
electrons are in, just as in the incompressible odd $q$ states discussed here
and in \cite{yang,moon}. (Note that this requires stronger inter-layer
correlations than in the 331 state supposed responsible for the observed
$\nu=1/2$ incompressible state \cite{nu1/2}.)
If such a state approximates the ground state when tunnelling $t$ is neglected,
then it is an itinerant ferromagnet
with fully polarized isospin, analogous to a Stoner-type metallic
ferromagnet, though the Coulomb interactions are, as
before, not fully SU(2) invariant, so the order can again be described simply
by an angle $\theta$. Bonesteel \cite{bonesteel} has discussed the opposite
case of a two layer system with the same correlations but without spontaneous
isospin polarization, in which
there are Fermi surfaces for both $\uparrow$ and $\downarrow$ spins. There
is also
the intermediate possibility of a partially polarized state at $t=0$.
All systems having a spontaneous polarization at $t=0$ have an
isospin stiffness but there are no compact,
well defined charged vortex excitations because the system has nonzero
compressibility for changes in $n_\uparrow+n_\downarrow$. Due to the stiffness,
a commensurate-incommensurate transition should occur in these cases also,
which might be observable through its effect on the Fermi surface \cite{HLR}.

To conclude, observation of the downward cusp in the activation energy gap
predicted here, as given in eq.\ (\ref{final}), on going through the
transition would confirm the general picture first presented in \cite{yang}.
However, the alternative possibility of significant disorder effects should
also be considered further.

I am grateful for discussions with J.P. Eisenstein, S.Q. Murphy, S.M. Girvin,
K. Yang, M.P.A. Fisher, D. Huse and S. Sachdev, and for the hospitality of
AT\&T Bell Labs. This work was supported by NSF-DMR-9157484.

\vspace*{\fill}

\begin{thebibliography}{99}
\bibitem{murph}
S.Q. Murphy, J.P. Eisenstein, G.S. Boebinger, L.N. Pfeiffer and K.W. West,
Phys.\ Rev.\ Lett.\ {\bf 72}, 728 (1994).
\bibitem{yang}
K. Yang, K. Moon, L. Zheng, A.H. MacDonald, S.M. Girvin, D. Yoshioka and S.-C.
Zhang, Phys.\ Rev.\ Lett.\ {\bf 72}, 732 (1994).
\bibitem{moon}
K. Moon et al, Indiana University preprint part I (1994), and in preparation.
\bibitem{wenzee}
X.-G. Wen and A. Zee, Phys.\ Rev.\ Lett.\ {\bf 69}, 1811 (1992);
Phys.\ Rev.\ B {\bf 47} 2265 (1993).
\bibitem{halpetc}
B.I. Halperin, Helv. Phys. Acta {\bf 56}, 75 (1983);
E.H. Rezayi and F.D.M. Haldane, Bull. Am. Phys. Soc {\bf 32}, 892 (1987).
\bibitem{leekane}
D.-H.Lee and C.L. Kane, Phys.\ Rev.\ Lett. {\bf 64}, 1313 (1990).
\bibitem{sondhi}
S. Sondhi, A. Karlhede, S.A. Kivelson and E.H. Rezayi, Phys.\ Rev.\ B
{\bf 47}, 16419 (1993).
\bibitem{ezawa}
Z.F. Ezawa and A. Iwazaki, Int. J. Mod Phys. B {\bf 19}, 3205 (1992);
Phys.\ Rev.\ B {\bf 47}, 7295 (1993).
\bibitem{HLR}
B.I. Halperin, P.A. Lee and N. Read, Phys.\ Rev.\ B {\bf 47}, 7312 (1993).
\bibitem{bak}
P. Bak, Rep. Prog. Phys. {\bf 45}, 587 (1982).
\bibitem{dennijs}
M. den Nijs, in {\it Phase Transitions and Critical Phenomena}, Vol 12,
eds C. Domb and J. Lebowitz (Academic Press, New York, 1988), pp. 219-333.
\bibitem{frank}
F.C. Frank and J.H. Van der Merwe, Proc. Roy. Soc. {\bf 198}, 205, 216 (1949).
\bibitem{poktal}
V. Pokrovsky and A.L. Talapov, Phys.\ Rev.\ Lett.\ {\bf 42}, 65 (1979).
\bibitem{fishfish}
M.E. Fisher and D.S. Fisher, Phys.\ Rev.\ B {\bf 25} 3192 (1982).
\bibitem{coppersmith}
S.N. Coppersmith et al, Phys.\ Rev.\ B {\bf 25}, 349 (1982).
\bibitem{yangpriv}
K. Yang, private communication.
\bibitem{callan}
A. Abouelsaood, C.G. Callan, C.R. Nappi and S.A. Yost, Nucl.\ Phys.\ B
{\bf 280}, 599 (1987).
\bibitem{fisher}
M.P.A. Fisher, private communication.
\bibitem{halpnewport}
B.I. Halperin, Surface Sci.\ {\bf 305}, 1 (1994).
\bibitem{nu1/2}
Y.W. Suen et al, Phys.\ Rev.\ Lett.\ {\bf 68}, 1379 (1992);
J.P.~Eisenstein et al, Phys.\ Rev.\ Lett.\ {\bf 68}, 1383 (1992).
\bibitem{bonesteel}
N.E. Bonesteel, Phys.\ Rev.\ B {\bf 48}, 11484 (1993).


\end{thebibliography}
\end{document}